\documentclass[twocolumn]{aastex631}

\usepackage{amsmath, amsfonts}
\usepackage{amssymb,amsmath,verbatim,mathtools,needspace,enumitem,etoolbox,graphicx,physics,microtype,afterpage,xspace,tabularx,lmodern,multirow,gensymb,placeins}

\newcommand{\nn}{\nonumber}
\newcommand{\be}{\begin{equation}}
\newcommand{\ee}{\end{equation}}
\def\be{\begin{equation}}
\def\ee{\end{equation}}
\newcommand{\beq}{\begin{eqnarray}}
\newcommand{\eeq}{\end{eqnarray}}
\usepackage{orcidlink}

\newcommand{\ciera}{\affiliation{Center for Interdisciplinary Exploration and Research in Astrophysics (CIERA),
Northwestern University, 1800 Sherman Ave, Evanston, IL 60201, USA}}
\newcommand{\dopNU}{\affiliation{Department of Physics and Astronomy, Northwestern University, 2145 Sheridan Road, Evanston, IL 60208, USA}}

\newcommand{\vl}{\ensuremath{\vec{\Lambda}}\xspace}
\newcommand{\vd}{\ensuremath{D}\xspace}
\newcommand{\vt}{\ensuremath{\vec{\theta}}\xspace}
\newcommand{\Msun}{\ensuremath{M_\odot}\xspace}

\newcommand{\alphaCE}{\ensuremath{\alpha_\text{\tiny CE}}}

\newcommand{\new}[1]{{\leavevmode{#1}}}

\begin{document}

\title{Dropping Anchor: Understanding the Populations of Binary Black Holes with Random and Aligned Spin Orientations}

\author[0000-0002-2536-7752]{Vishal Baibhav} 
\email{vishal.baibhav@northwestern.edu}
\ciera
\author[0000-0002-2077-4914]{Zoheyr Doctor} 
\ciera
\author[0000-0001-9236-5469]{Vicky Kalogera }
\ciera\dopNU

\begin{abstract}
The relative spin orientations of black holes (BHs) in binaries encode their evolutionary history:
BHs assembled dynamically should have isotropically distributed spins, while spins of the BHs originating in the field should be aligned with the orbital angular momentum. 
In this article, we introduce a simple population model for these dynamical and field binaries that uses spin orientations as an anchor to disentangle these two evolutionary channels.
We then analyze binary BH mergers in the Third Gravitational-Wave Transient Catalog (GWTC-3) and ask whether BHs from the isotropic-spin population possess different distributions of mass-ratio, spin magnitudes, or redshifts from the preferentially-aligned-spin population. 
We find no compelling evidence that binary BHs in GWTC-3 have different source-property distributions depending on their spin alignment, but we do find that the dynamical and field channels cannot both have mass-ratio distributions that strongly favor equal masses. 
We give an example of how this can be used to provide insights into the various processes that drive these BHs to merge. We also find that the current detections are insufficient in extracting differences in spin magnitude or redshift distributions of isotropic and aligned spin populations. 
\end{abstract}

\section{Introduction}

Gravitational waves offer an unprecedented look into some of the most rare phenomena in the universe, in particular the inspirals and mergers of pairs of compact objects. The properties of these merging black holes (BHs) and neutron stars such as their masses, spins, and redshifts provide clues on how these mergers come to occur. So far, nearly 100 gravitational waves have been detected by the Advanced Laser Interferometer Gravitational-Wave Observatory (LIGO) \citep{LIGOScientific:2014pky} and Advanced Virgo \citep{VIRGO:2014yos} detectors. These detections have been reported on by the LIGO-Virgo-KAGRA Scientific Collaboration (LVK) \citep{LIGOScientific:2018mvr,LIGOScientific:2020ibl,LIGOScientific:2021djp} as well as by other teams \citep[e.g.][]{Nitz:2021zwj, Olsen:2022pin}. 

So far, gravitational waves (GWs) have been seen from mergers of two BHs, two neutron stars, and neutron stars with BHs. The individual GW signals have shown a range of progenitor masses, mass ratios, spins, and redshifts \citep{LIGOScientific:2016vbw,LIGOScientific:2020stg,LIGOScientific:2020zkf}, and one even produced a display of electromagentic counterparts that were observed by other facilities \citep{LIGOScientific:2017vwq,LIGOScientific:2017ync}. Beyond inferences on individual systems, the full catalog of gravitational waves can be leveraged to measure the {\it population distributions} of underlying merger properties in the universe \citep[e.g.][]{Mandel:2018mve}. These population-level inferences offer even more clues into the histories of merging compact objects. For example, the distribution of black-hole masses of the more-massive merger components $m_1$ peaks around 5-10 $\Msun$ and has an additional bump around 35 $\Msun$, and there is a preference for the secondary masses $m_2$ to be near their respective primary's masses \citep{LIGOScientific:2020kqk,Tiwari:2020otp,Tiwari:2021yvr,Edelman:2021zkw,LIGOScientific:2021psn,Sadiq:2021fin}. In terms of spins, the distribution of effective inspiral spin $\chi_{\rm eff}$, the mass-weighted projection of component spin vectors onto the orbital angular momentum, peaks near zero, but is definitively biased towards positive values \citep{LIGOScientific:2020kqk,LIGOScientific:2021psn}. Furthermore, $\chi_{\rm eff}$ appears to be anti-correlated with the binary mass ratio $q\equiv m_2/m_1$ \citep{Callister:2021fpo,LIGOScientific:2021psn,Adamcewicz:2022hce} 
 with a possible correlation with redshift \citep{Biscoveanu:2022qac} \new{and mass \citep{Franciolini:2022iaa}}.  Rather than modeling effective inspiral spins, one can also model the spins of individual BHs and their spin directions. The individual BH dimensionless spins tend to be small $\sim 0.2$, but the tilt angle distributions are not particularly well-measured when accounting for systematic uncertainties \citep{Galaudage:2021rkt,Callister:2022qwb,Vitale:2022dpa,Edelman:2022ydv,Golomb:2022bon}. Ultimately, any proposed formation pathways of the detected BH mergers must agree with these empirical findings. 

The two most well-studied formation channels are isolated binary evolution in the field and dynamical interactions in clusters (see e.g.~\citep{Mandel:2018hfr,Mapelli:2018uds} for reviews).  For isolated binaries, binary interactions such as stable mass transfer and CE phases may aid in the formation of two BHs that can merge within a Hubble time, though the exact efficacy of each of these processes in producing BH mergers is still under debate (see \citet{Gallegos-Garcia:2021hti} and references therein). Alternatively, dynamical channels predict that binary BHs (BBHs) form and harden through three-body encounters in dense stellar clusters. {Other scenarios for the formation and merger of BBHs include chemically homogenous evolution of isolated binaries~\citep{Marchant:2016wow,deMink:2016vkw}, AGN disks~\citep{Stone:2016wzz,Bartos:2016dgn,Leigh:2017wff}, secular interactions in triples~\citep{Silsbee:2016djf,Hoang:2017fvh,Fragione:2018yrb}, and primordial BHs~\citep{Bird:2016dcv}.} Different formation pathways leave different imprints on the properties of the BBH population, including the binary masses, spins, eccentricities, and redshift evolution. Measuring these distributions informs us on the environment in which BBHs form and evolve~\citep{Zevin:2017evb,Taylor:2018iat,Wysocki:2018mpo, Roulet:2018jbe,LIGOScientific:2018jsj}.

One of the most promising signatures is the distribution of BH spin orientations: systems formed through dynamical interactions are expected to have isotropic spin orientations, whereas binaries born in the field are more likely to have spins aligned with the orbital angular momentum~\citep{Gerosa:2013laa,Vitale:2015tea,Rodriguez:2016vmx,Farr:2017gtv,Gerosa:2018wbw,Stevenson:2017dlk}. Using GW observations, we can separate populations of isotropic and aligned binaries.
In this work, we ask the question: Can we observe differences between the population of BHs with isotropic spins and the population with aligned spins? If the properties of systems from the isotropically-distributed spin directions are different from aligned-spin systems, we could use that as an anchor to study the evolutionary histories of BHs coming from cluster and field formation channels.

This work is organized as follows: we first introduce a generic mixture-model framework for separating isolated and cluster binary BHs, using spin tilts as an anchor. We then apply this framework to investigate whether the mass ratio distributions are different between these two formation channels. Section~\ref{sec:astro} illustrates how these results can be compared back to population synthesis studies. We conclude in Section~\ref{sec:conclusions} with some remarks on how the framework described herein can be employed.

\section{Mixture models}
\label{sec:mixturemodels}

We employ a hierarchical Bayesian inference framework to measure  distributions of properties of individual BBH mergers  (such as masses, spins, redshifts, etc.) using only gravitational-wave data. The distribution of the individual binary properties, \vt can be parameterized in terms of {\em unknown} population-level hyper-parameters \vl. We wish to  infer this \vl given a catalog of GW detections. We elaborate on the hierarchical Bayesian inference framework in Sec.~\ref{sec:HBA}.

To model potential differences in populations of field and cluster BH-merger properties, we divide the BBH population into two sub-populations, one requiring isotropic spins and another with preferentially aligned spins, respectively. We further separate the binary parameters into two sets:
\be
\vt=\{\vt_{\rm mix},  \vt_{\rm pure}\}\, ,\nn
\ee
where $\vt_{\rm mix}$ come from a mixture of the two subpopulations, and $ \vt_{\rm pure}$ are drawn from a common distribution for both subpopulations. Similarly, the hyperparameters corresponding to  $\{\vt_{\rm mix}$ and  $\vt_{\rm pure}\}$ can also be divided into 
\be
\vl=\{\vl_{\rm mix}, \vl_{\rm pure}\}\,.\nn
\ee
$\vl_{\rm mix}$ further consists of
\be
\vl_{\rm mix}=\{\vl_I, \vl_A\}
\ee
where $\vl_I$ is the set of hyperparameters that parameterize the binary parameters $\vt_{\rm mix}$ of the isotropic subpopulation, while $\vl_A$ parameterizes $\vt_{\rm mix}$ of the aligned subpopulation.

One approach to differentiate the aligned and isotropic binaries is to look at their spin directions directly:
\be
\cos\theta_i = \hat{\mathbf{S_i}}\cdot\hat{\mathbf{L}}
\ee
where $\theta_i$ ($=1,2$) is the angle between the BH spin $\mathbf{S_i}$, and the orbital angular momentum $\mathbf{L}$. Extending the model introduced in \citet{Talbot:2017yur}, we assume these tilts are identically distributed as a mixture between an isotropic component and a preferentially-aligned component:
\be
\begin{aligned}
&p_{\rm mix}(\vt_{\rm mix}, \cos\theta_i|\vl_{\rm mix},\zeta,\sigma_t) = \frac{(1-\zeta)}{4} \; p(\vt_{\rm mix} | \vl_{I})\\
&\hspace{2cm} + \zeta \;p(\vt_{\rm mix} | \vl_{A}) \; p(\cos\theta | \sigma_t)\,.
\end{aligned}
\label{eq:mix_pop}
\ee
Here, $\zeta$ is the mixing fraction of events comprising the aligned subpopulation, and the remaining fraction $1-\zeta$ is assigned to the isotropic-spin population. The spin directions of aligned binaries are modeled as a truncated-Gaussian centered at $\cos\theta=1$
\be
p(\cos\theta | \sigma_t)=\prod_{i=1,2}{\cal {N}}_{[-1,1]}(\cos\theta_i|1,\sigma_t)
\ee  

Unfortunately, individual properties of BH spin (such as magnitudes and directions) can not be measured very accurately with GWs. Instead the most well-measured spin parameter is the effective inspiral spin
\be
\chi_{\rm eff} = \frac{\chi_1 \cos\theta_1+ q \chi_2 \cos\theta_2}{1+q}
\ee
which is the leading order spin contribution and a constant at 2PN level~\citep{Damour:2001tu,Racine:2008qv}. Here $q=m_2/m_1$ is the ratio of BH masses (where $m_1$ is the heavier BH while $m_2$ is the lighter BH); $\chi_1$ and  $\chi_2$ are the spins of primary and secondary BHs respectively. $\chi_{\rm eff}$ has long been proposed as a tool to differentiate between binaries formed in the isolated channel and those assembled dynamically. Since orientations of BHs formed in stellar clusters are isotropically distributed, this leads to a symmetric $\chi_{\rm eff}$ distribution centered at $\chi_{\rm eff}=0$, while the field binaries should be preferentially aligned~\citep{Kalogera:1999tq}, leading to a distribution skewed towards positive values of $\chi_{\rm eff}$~\citep{Rodriguez:2016vmx,Farr:2017gtv,Wysocki:2017isg,Gerosa:2018wbw,Ng:2018neg}. For this reason, another approach to differentiate population parameters of the aligned and isotropic binaries is to model this mixture as
\be
\begin{aligned}
&p_{\rm mix}(\vt_{\rm mix}, \chi_{\rm eff}|\vl_{\rm mix},\zeta) =\\
&\hspace{1cm}\zeta \;p(\vec{\theta}_{\rm mix} | \vl_{A}) \; N_{[-1,1]}(\chi_{\rm eff} | \mu_{\rm eff, A}, \sigma_{\rm eff, A})\\
&\hspace{1cm} +(1-\zeta)\; p(\vec{\theta}_{\rm mix} | \vl_{I})  N_{[-1,1]}(\chi_{\rm eff} | \mu_{\rm eff, I}=0, \sigma_{\rm eff, I})
\end{aligned}
\label{eq:mix_pop2}
\ee
where the first term models the distribution of $\chi_{\rm eff}$ for aligned spin population  as a gaussian truncated between $[-1,1]$ with $\mu_{\rm eff, A}>0$, while the second term represents the isotropic population with a truncated gaussian centered around $\mu_{\rm eff, I}=0$.

In this study, we will focus on using the spin directions $\cos\theta_{1, 2}$ as an indicator of isotropic and aligned populations. We will assume that priors on $\zeta$ are uniform between $0$ and $1$, while priors on $\sigma_t$ are uniform between $0.1$ and $4$.
Our full population prior, $p(\vt | \vl)$, is the product of the mixture [Eqs.~\ref{eq:mix_pop}] and pure population models:
\be
p(\vt | \vl)=p_{\rm mix}(\vt_{\rm mix}, \cos\theta_i|\vl_{\rm mix},\zeta,\sigma_t)\times p(\vt_{\rm pure}|\vl_{\rm pure})
\ee

In this study, we will often parameterize some of the binary properties \vt (either pure $\vt_{\rm pure}$ or in the mixture of aligned and isotropic components $\vt_{\rm mix}$). The parameters not studied will be assumed to come from distributions with hyperparameters fixed to their median values as inferred from  \citet{LIGOScientific:2021psn}.

\section{Do isotropic and aligned binaries pair differently?}
\label{sec:q-dist}

\begin{figure*}[htbp!]
    \includegraphics[width=\columnwidth]{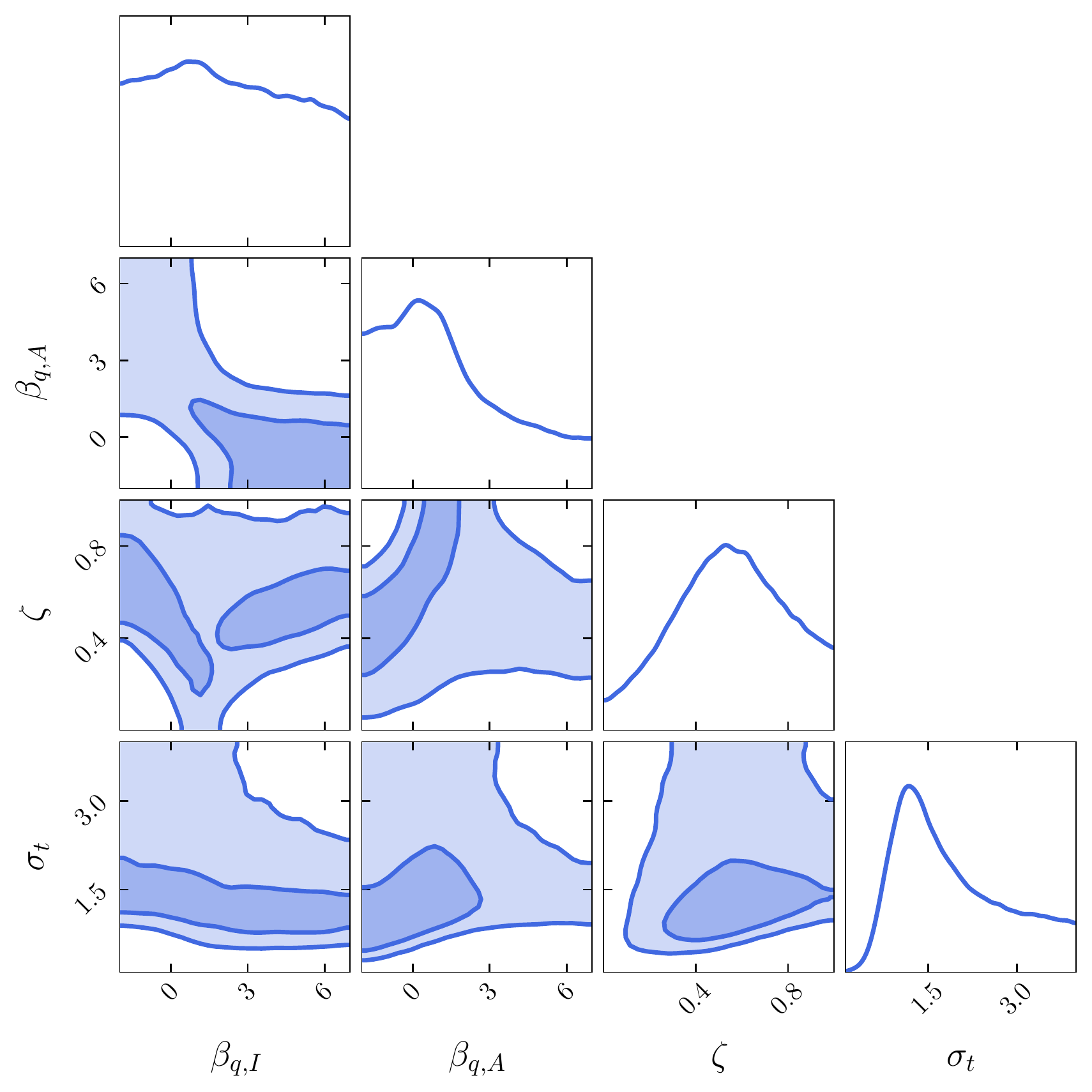} \includegraphics[width=\columnwidth]{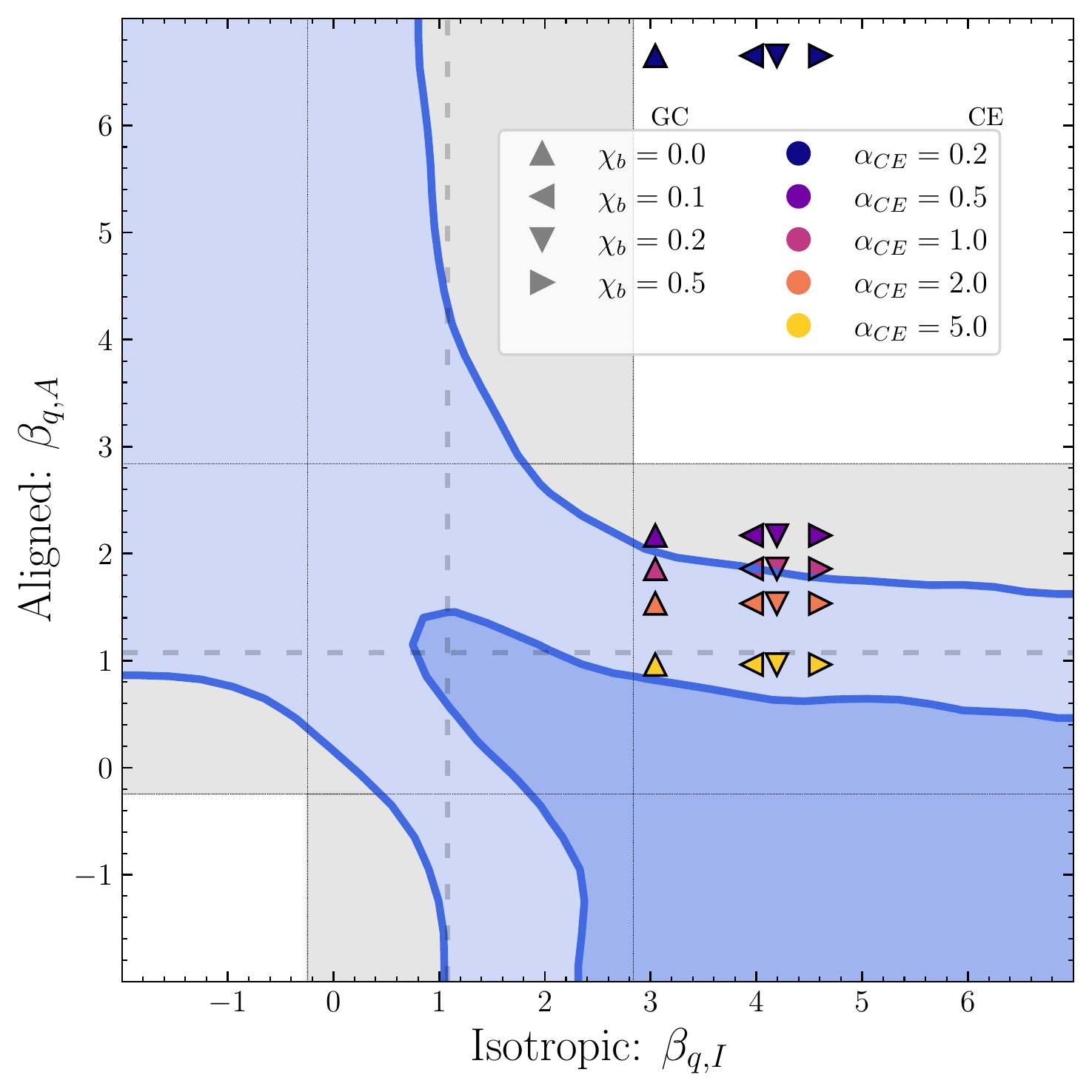}
\caption{Left panel: posteriors of all the hyperparameters in the \texttt{q-mix} model: $\beta_{q,I}$, $\beta_{q, A}$, $\zeta$ and $\sigma_t$. We plot the $68\%$ and $90\%$ intervals. Right: we plot the posteriors of $(\beta_{q,I},\,\beta_{q, A})$. Grey shaded region marks the $90\%$ credible regions on $\beta_q$ recovered in GWTC-3~\citep{LIGOScientific:2021psn}. We also show the $\beta_{q, I} = \beta_{q*}$ extracted using the globular cluster models in~\citet{Zevin:2020gbd} for different values of BH natal spins $\chi_b$ (shown by markers of different shape). Similarly, we use the CE models to represent aligned binaries with the $\beta_{q, A} = \beta_{q*}$ extracted  for different values of CE efficiencies \alphaCE (shown by markers of different colors).
}
    \label{fig:beta_IA}
\end{figure*}

We use the mixture model describe in Sec.~\ref{sec:mixturemodels} to discern if isotropic and aligned populations have mass pairing. For simplicity, we assume that the primary-mass distribution of both isotropic and aligned binaries follow the \texttt{PowerLaw+Peak} model described in \citep{LIGOScientific:2021psn} that is the distribution of primary masses $m_1$ is parameterized as the mixture model of a power law and a gaussian distribution~\citep{Talbot:2018cva}.
The primary mass model consists of $7$ parameters: power law slope $\alpha$, mixture fraction $\lambda_{\rm peak}$, minimum ($m_{\min}$) and maximum BH mass ($m_{\max}$),  mean ($\mu_m$) and standard deviation ($\sigma_m$) of the gaussian component and a smoothing factor ($\delta_m$).
With $\vl_m=\{\alpha, m_{\min}, m_{\max}, \delta_m, \mu_m, \sigma_m, \lambda_{\rm peak}\}$
\be
p(m_1|\vl_m) = (1-\lambda_{\rm peak}){\mathcal P} + \lambda_{\rm peak} \, {\mathcal G}
\label{eq:p_m1}
\ee
where
\be
{\mathcal P} \propto m_{1}^{-\alpha} S(m_1,  m_{\min}, \delta_m) \Theta(m_{\max}-m_1)
\ee
is the power-law distribution and
\be
{\mathcal G}\propto \exp\left(-\frac{(m_1-\mu_m)^2}{2\sigma_m^2}\right)S(m_1,  m_{\min}, \delta m).
\ee
is a Gaussian peak. In addition, this model employs a smoothing function at the low-masses, $S(m,  m_{\min}, \delta m)$, which rises from zero to one as mass increases from $ m_{\min}$ to $ m_{\min}+\delta_m$,
\begin{align}
S(m,  m_{\min}, \delta_m) &= \frac{1}{1+e^{f(m- m_{\min}, \delta_m) }},\nn\\
f(m, \delta_m) &= \frac{\delta_m}{m} - \frac{\delta_m}{m - \delta_m}.
\end{align}

Since we are only focusing on discerning whether binaries with aligned and isotropic spins have different mass-ratio distributions, we fix the hyperparameters governing the primary mass distribution (that is $\vl_m=\{\alpha, m_{\min}, m_{\max}, \delta_m, \mu_m, \sigma_m, \lambda_{\rm peak}\}$) to the median values obtained in \citet{LIGOScientific:2021psn}. We allow the aligned and isotropic binaries to have different mass ratio distributions, that is, $\vt_{\rm mix}={q}$ governed by hyperparameters $\vl_{\rm mix}=\{\beta_{q, I}, \beta_{q, A}\}$. Here $\beta_{q, I}$ is the power-law slope of the $q$ distribution of isotropic-spin binaries while $\beta_{q, A}$ is the corresponding power-law slope of aligned-spin binaries (allowing for a smooth turn-on for low-mass secondary BHs),
\be
p(q|m_1, \beta_{q, i}, \delta_m,  m_{\min}) \propto q^{\beta_{q, i}} S(m_2, m_{\min},\delta m) \Theta(m_{1}-m_{2}).
\label{eq:pq}
\ee
where $i=I, A$. We call this model \texttt{q-mix}. We assume priors on $\beta_{q, I}$ and $\beta_{q, A}$ are uniform between $-2$ and $7$.

Fig.~\ref{fig:beta_IA} shows results when analyzing GWTC-3 binary BHs with the \texttt{q-mix} model. The left panel shows the $68\%$ and $90\%$ credible regions for the recovered power-law slopes of $q$ distribution for isotropic and aligned populations ($\beta_{q,I}, \beta_{q,A}$), the fraction of aligned binaries ($\zeta$) and standard deviation of aligned-spin tilts $\sigma_t$. Most notably, part of $\beta_{q,I} = \beta_{q,A}$ line is included in the $90\%$ credible region, implying that there is no strong evidence for different $\beta_{q}$ between the isotropic-spin and aligned-spin populations. Furthermore, the Bayes factor in favor of different $\beta_q$'s is only $1.62$. Nevertheless, we can probe the rest of the posterior to understand what alternative hypotheses may still be viable given the data. Firstly, we observe that $\beta_{q, I}$ and $\beta_{q, A}$ are possibly anti-correlated. Consequently, two regions in $(\beta_{q,I},\, \beta_{q,A})$ parameter space are disfavored: i) both $\beta_{q,I}$ and $\beta_{q, A}$ being small ($\lesssim 1$), ii) both $\beta_{q,I}$ and $\beta_{q, A}$ being large ($\gtrsim 2$). This implies that both isotropic and aligned binaries cannot have an extremely selective pairing or random pairing, which comports with the inferred $\beta_q \sim 1$ assuming a single mass-ratio distribution for all BBHs \citep{LIGOScientific:2021psn}. Interestingly, there is some support for $\beta_{q, I}$ and $\beta_{q, A}$ having opposite signs. In that case, equal mass-ratio events would be dominated by one channel, while unequal mass events would be dominated by the other.  Most of the posterior samples have dissimilar $\beta_{q,I}$ and $\beta_{q,A}$.
In addition, the posteriors on $\beta_{q, I}$ are least informative when $\beta_{q,A}$ is between $0$--$2$. A similar statement can be made for $\beta_{q, A}$ when $\beta_{q, I}\simeq 1$. 

Our results also have implications for the fraction of aligned systems $\zeta$.  If $\beta_{q, I}\simeq 1$, then the data slightly prefer that the majority of binaries have random spin orientations, but a range of mixing fractions are possible. On the other hand, if $\beta_{q, A} \simeq 1$, there is likely a larger contribution from the aligned-spin population.
In addition, we can exclude $\zeta=0$ if $\beta_{q,I}<0$  or $\beta_{q, I}\gtrsim 3$. This is because if isotropic-spin binaries have unequal masses (equal masses), the aligned spin channel must be invoked to explain  the mergers involving  similar masses  (unequal masses). For a similar reason, $\zeta=1$ is disfavored for $\beta_{q, A}<0$  or $\beta_{q, A}\gtrsim 3$.

There is a positive correlation between $\zeta$ and $\sigma_t$ as also reported in~\citet{LIGOScientific:2020kqk}: if the tilts of aligned binaries are small, then their fraction is also small, and most of the BBHs must be isotropic to explain the observations. If we increase $\sigma_t$, the aligned binaries are allowed to have larger tilts, and binaries that were earlier considered isotropic are now considered aligned. A majority of the samples identified as aligned have very large tilts.

As mentioned earlier, BBHs detected during O3 show evidence of anti-correlation between the mass ratios and spins, with equal masses possessing smaller $\chi_{\rm eff}$ and unequal-mass mergers exhibiting  larger $\chi_{\rm eff}$ \citep{Callister:2021fpo,LIGOScientific:2021psn}.  There is a possibility that this behavior can be explained using by assuming isotropic and aligned binaries have different $q$ distributions. In particular, if isotropic binaries have a distribution that strongly favors $q=1$ (that is large $\beta_{q, I}$) while the aligned binaries dominate at smaller mass ratios (either small $\beta_{q, A}$ or $\beta_{q, A}\leq 0$). In this case, the $q=1$ region would be populated with isotropic binaries with $\chi_{\rm eff}$ symmetric around $0$, while the aligned population will possess  smaller mass ratios and larger values of $\chi_{\rm eff}$. So as we move from $q=1$ to smaller $q$, the mean of the $\chi_{\rm eff}$ distribution increases as the presence of aligned binary increases, thereby explaining the $q-\chi_{\rm eff}$ correlation. Another component that controls the $q-\chi_{\rm eff}$ correlation is the spin magnitude of individual BHs. We defer the discussion on the spin magnitude of isotropic and aligned systems to Appendix~\ref{sec:sz}.

In this section, we allowed the isotropic and aligned populations to have different $q$ distributions, however, this could also be extended to include primary masses ($m_1$) as well, i.e., $\vt_{\rm mix}=\{ m_1, q\}$. In this study, we do not evaluate the differences in the primary-mass distributions of the two subpopulations because the model becomes complicated with $16$ hyperparameters ($8$ for each subpopulation), and defer such investigations to future work with a larger catalog of sources.

\begin{figure}[t]
    \includegraphics[width=\columnwidth]{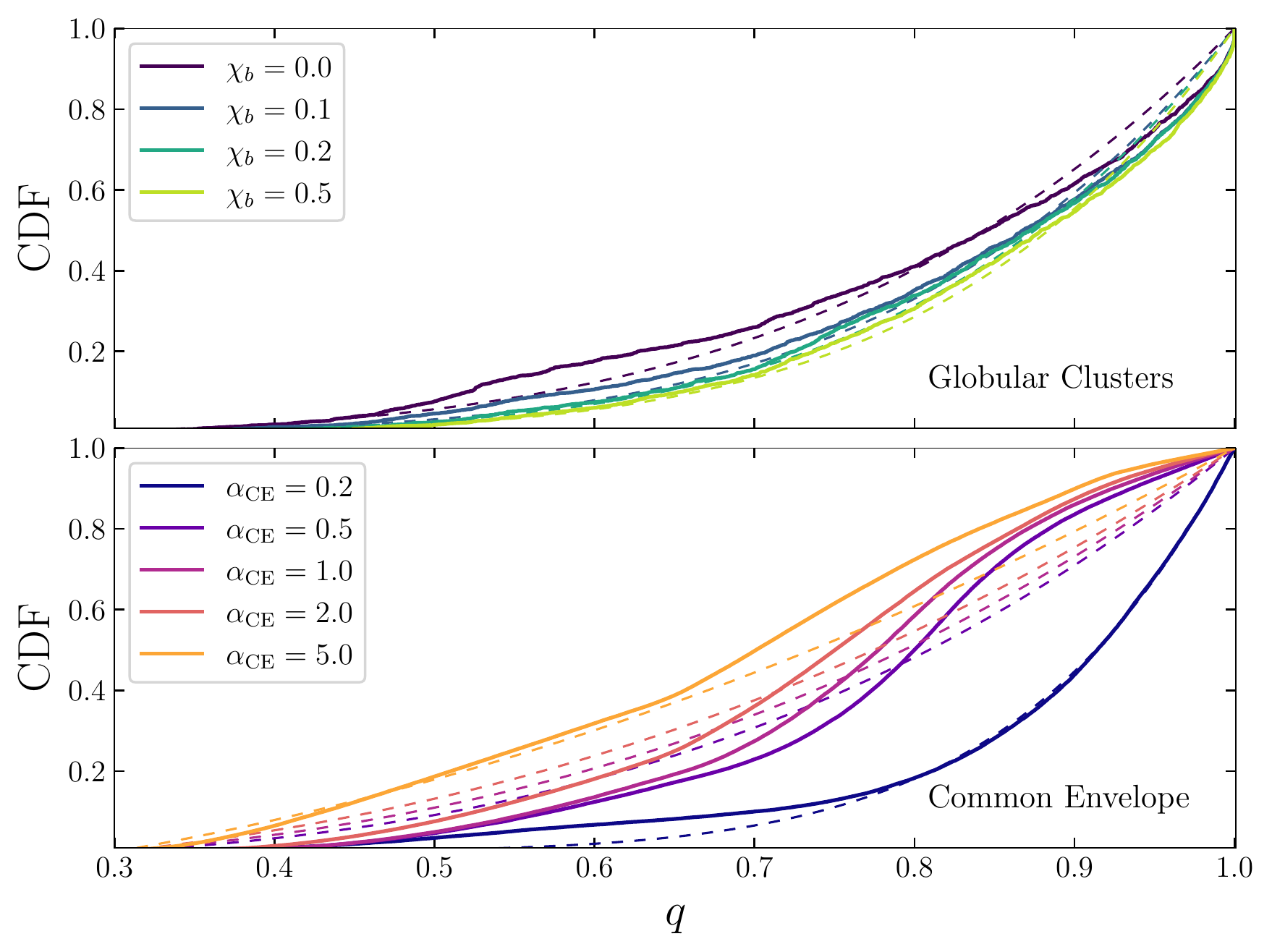}
    \caption{Cumulative density function of $q$ distribution of BBHs in the dynamical (top) and isolated (bottom) scenario. In the top panel (bottom) we show the CDFs for varying $\chi_b$ (\alphaCE). The dashed lines show the power-law fit ($p(q)\propto q^{\beta_{q*}}$) to these astrophysical distributions.}
    \label{fig:beta_astro}
\end{figure}
    
\begin{figure}[t]
    \includegraphics[width=\columnwidth]{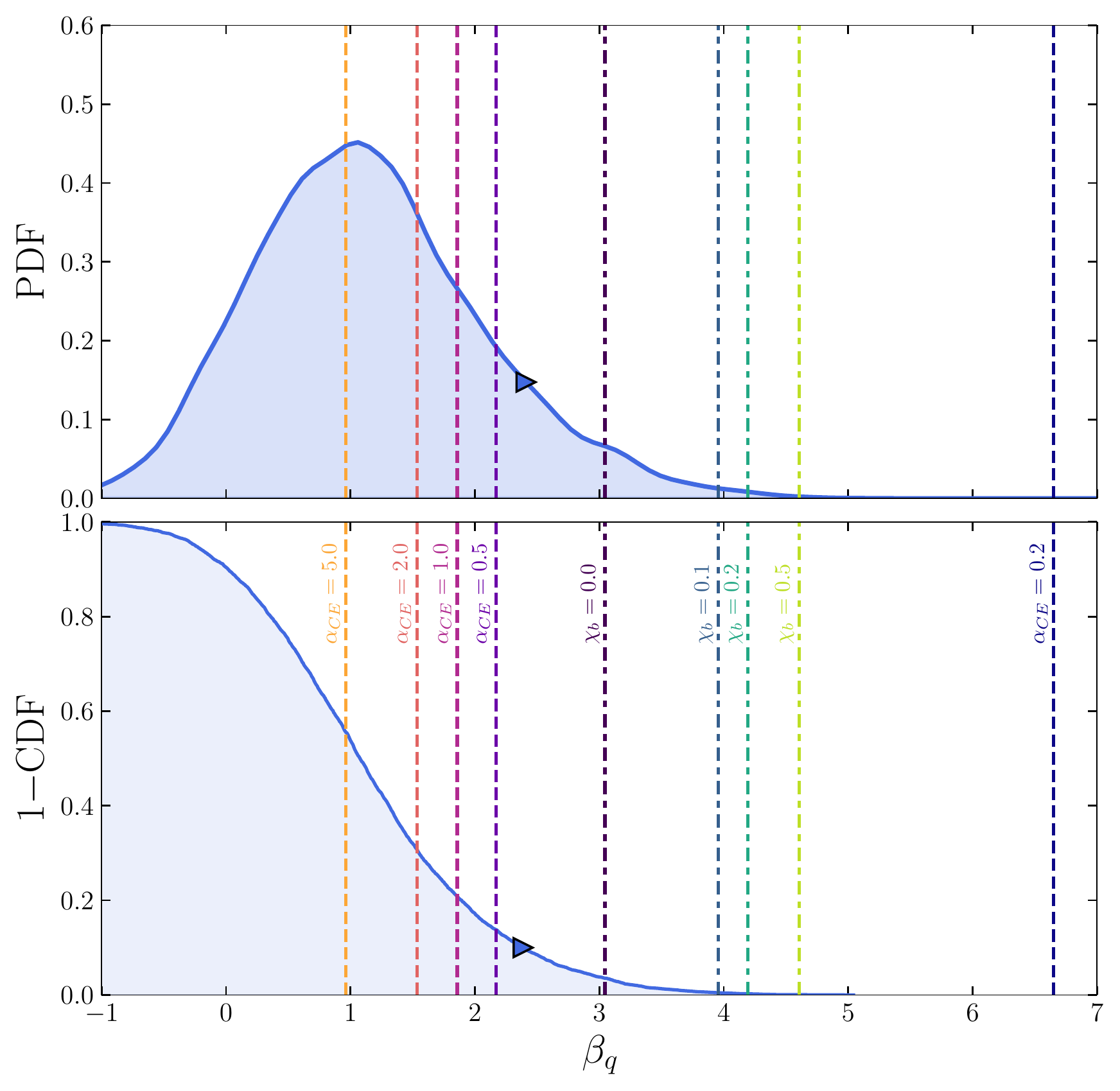}
    \caption{The distribution of $\beta_q$ for the LVK \texttt{PowerLaw+Peak} model. Vertical lines mark the $\beta_{q*}$ obtained by fitting astrophysical models: dot-dashed lines represent the pairing functions of GC models for $\chi_b \in \{ 0, 0.1, 0.2, 0.5\}$ with smaller $\chi_b$ yielding smaller $\beta_{q*}$, while the dotted lines represent the pairing functions of CE models for $\alphaCE \in \{ 0.2, 0.5, 1.0, 2.0,  5.0 \}$ with smaller \alphaCE~yielding larger $\beta_{q*}$. The triangle marker represents the $90\%$ upper limit of the $\beta_q$ distribution.}
    \label{fig:beta_LVK}
\end{figure}

\begin{figure*}[t]
    \includegraphics[width=\columnwidth]{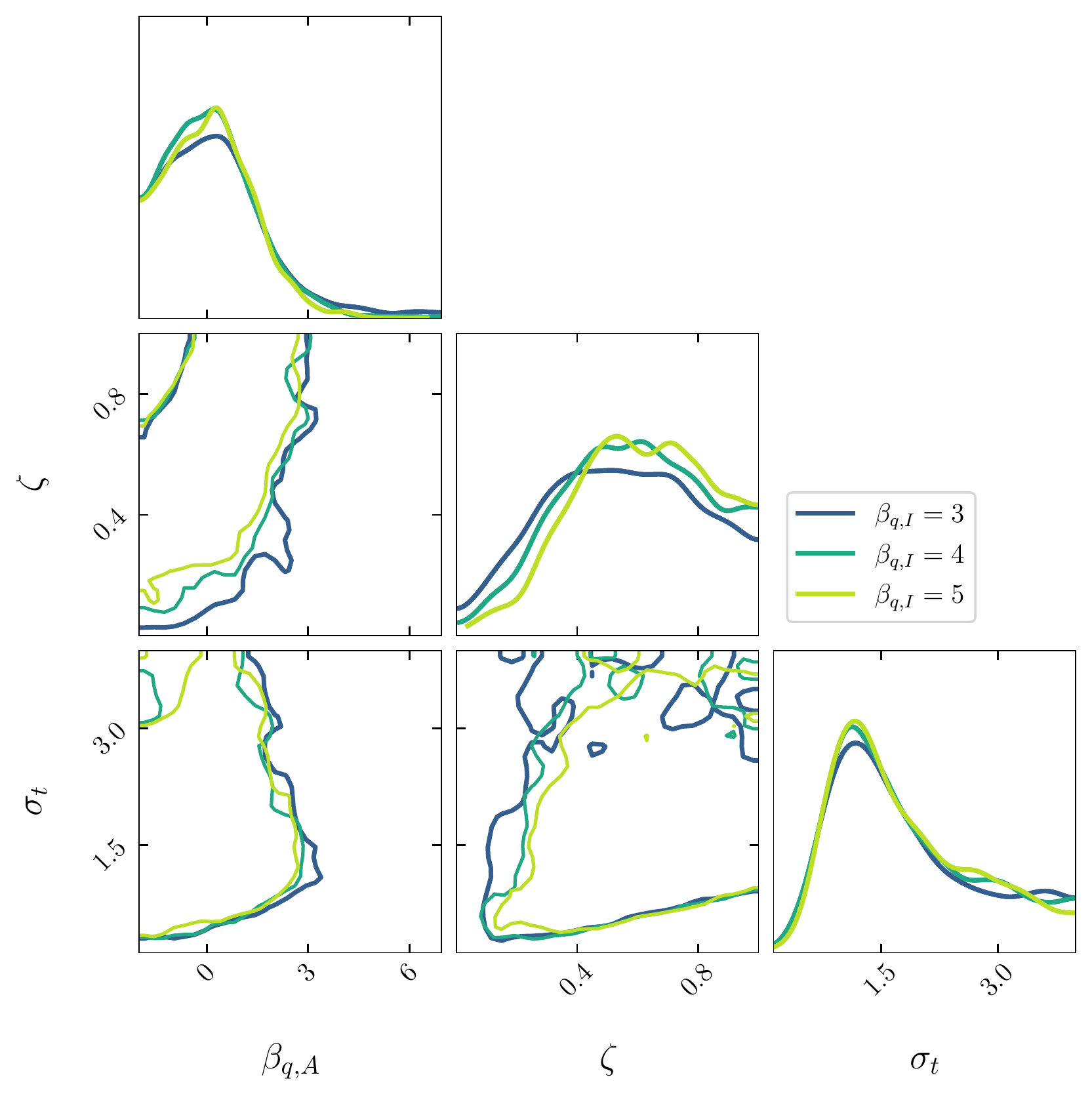} \includegraphics[width=\columnwidth]{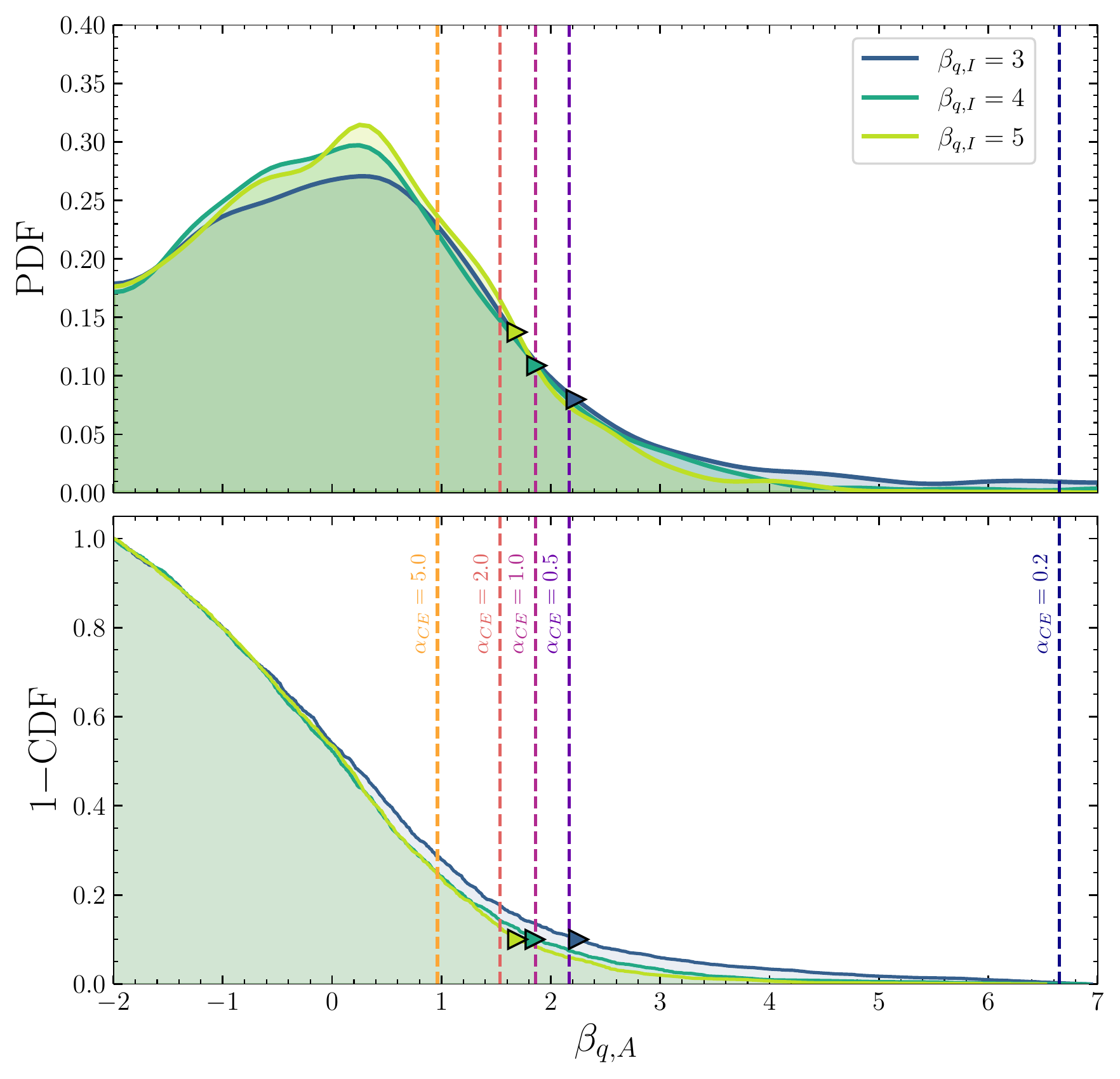}
\caption{Left panel: posteriors of all the hyperparameters in the \texttt{q-mix} model when $\beta_{q,I}=3, 4$ and $5$: $\beta_{q, A}$, $\zeta$ and $\sigma_t$. We plot the $68\%$ and $90\%$ intervals.  Right: the distribution of $\beta_{q, A}$ at the same fixed $\beta_{q,I}$ as the left panel. Vertical lines mark the $\beta_{q*}$ obtained by fitting CE models for $\alphaCE \in \{ 0.2, 0.5, 1.0, 2.0,  5.0 \}$ with smaller \alphaCE~yielding larger $\beta_{q*}$. The triangle marker represents the $90\%$ upper limit of the $\beta_{q, A}$ distribution.}
    \label{fig:betaI_fixed}
\end{figure*}

\subsection{Astrophysical implications - an illustration}
\label{sec:astro}
One of the long-standing problems in GW astrophysics is the origin of merging compact-object binaries and processes that drive mergers. Multiple channels have been proposed to explain such mergers, some of which could contribute to the detection of BBHs by LVK. However, categorizing these detections by the formation pathways is a challenging task, and to complicate matters, these pathways are plagued by theoretical uncertainties. The task's difficulty is greatly alleviated by the tell-tale signatures exhibited by only a subset of the proposed formation channels. Since the broad differences in spin directions are immune to astrophysical uncertainties, we use a framework to employ the spin directions as one such signature to separate multiple sub-populations. Studying other properties (for example, the mass-pairing function discussed above) of these sub-populations can further allow us to get a handle on different unknowns that plague their formation mechanisms.

This section gives an example of how the framework developed here can be applied to constrain and compare various astrophysical processes across different formation channels. For illustration purposes, we will use one of the isolated and dynamical channels from \citet{Zevin:2020gbd}.
Among the three field formation scenarios in \citet{Zevin:2020gbd}, we use the late-phase common envelope (CE) to represent the isolated channel\footnote{Note that~\citet{Gallegos-Garcia:2021hti} finds that common-envelope phases are unlikely to lead to BBH mergers, but we nevertheless make use of common-envelope models for illustration, since detailed astrophysical modeling in not in the scope of this article.}.
For this channel, the POSYDON framework~\citep{2022arXiv220205892F} was used to combine detailed MESA binary simulations with the COSMIC population synthesis code~\citep{Breivik:2019lmt}. The mass ratio distribution of this channel is only governed by the CE efficiency $\alphaCE \in [0.2, 0.5, 1.0, 2.0, 5.0]$, with large \alphaCE~leading to efficient CE evolution. Small \alphaCE ($\sim 0.2$) tends to show a preference for BBHs with similar masses. For  $\alphaCE>0.2$, we find that the $q$-distribution peaks around $\sim 0.8$, with larger \alphaCE~producing more unequal mass mergers. We also plot the cumulative density function for different \alphaCE ~in Fig.~\ref{fig:beta_astro}. The mass ratio distribution of the common-envelope channel 
used in \citet{Zevin:2020gbd} does not depend on the natal spin of BHs $\chi_b$.

For the dynamical channel, we consider BBH mergers in old, metal-poor globular clusters~(GC). The GC models are taken from a grid of 96 simulated clusters using the cluster Monte Carlo code CMC~\citep{Rodriguez:2019huv}: 24 models (with a range of initial cluster masses, metallicities, and half-mass radii) with 4 different natal BH spins ($\chi_b \in [0, 0.1, 0.2, 0.5]$). The mass ratio distribution of BBHs originating in the GCs depends on $\chi_b$. This dependence is primarily due to hierarchical mergers. BHs born from stellar collapse (henceforth 1g BHs) have a similar $q$-distribution for all $\chi_b$. BHs with small 1g spins impart very small GW recoils to the remnant BHs (henceforth 2g BHs), which can be retained inside the cluster and likely merge again with 1g BHs. These 1g+2g mergers have mass ratios peaking at $\sim0.5$~\citep{Rodriguez:2019huv,Kimball:2020qyd,Kimball:2020opk,Kritos:2022ggc}. As the BH natal spin $\chi_b$ increases, GW recoils received by the remnants increase drastically, making their retention difficult. This significantly reduces the number of BBHs mergers near $q\sim 0.5$ in GCs. We also plot the cumulative density function for different $\chi_b$ in Fig.~\ref{fig:beta_astro}.

Since, in this study, we assume that the mass ratios of both isotropic and aligned sub-populations are distributed as a power law, we fit the cumulative density of mass ratios of astrophysical populations used in \citet{Zevin:2020gbd} as 
\be
p(q^*<q) = \frac{q^{\beta_{q*} +1}-q_{\rm min}^{\beta_{q*} +1}}{1-q_{\rm min}^{\beta_{q*} +1}}
\ee
where $q_{\rm min}$ is the minimum mass ratio in the simulation and $\beta_{q*}$ is the power law slope of the probability density function.
We plot their cumulative in Fig.~\ref{fig:beta_astro}, and we use the $\beta_{q*}$ as a proxy for mass-pairing resulting from the CE phase in the field and dynamics in globular clusters. Here, we stress that, in practice,  power laws are not ideal for fitting most of the astrophysical distributions. This is especially true if a particular channel also contains multiple sub-populations. For example, in GCs, the power law fails to appropriately  fit the 1g+2g mergers, which dominate at $q\sim 0.5$. This is important when natal spins are small ($\chi_b\simeq 0$) as up to $15\%$ of all GC mergers contain a 2g BH. In addition, the power laws also fail to fit appropriately when the q-distribution peaks away from $1$. This is the case for all common-envelope models (except $\alphaCE=0.2$) used in this study. Nonetheless, for illustration, we assume that $q$-distributions in these astrophysical models follow a power law.

We can first compare the mass pairing in the astrophysical models with the results from \texttt{PowerLaw+Peak} model when $q$ distributions are not separated based on the distribution of their spin directions
\citet{LIGOScientific:2021psn} also measured the pairing of BBHs $\beta_q=$ with a 90\% upper limit of $2.38$. This excludes astrophysical scenarios where all contributing formation channels produce extremely selective pairing with $q=1$. In Fig.~\ref{fig:beta_LVK}, we plot the recovered $\beta_q$ from the LVK analysis, along with $\beta_{q*}$ extracted from the aforementioned individual astrophysical models. Under the strong assumption that only one channel contributes to the BBH population, we find that the only models that are not disfavored are common-envelope models with $\alphaCE\ge 0.5$. On the other hand, all the globular cluster models and CEs models with $\alphaCE=0.2$ are unable to fit the data with as the sole formation channel.

Of course, there is no reason for only one formation channel to contribute to the BBH population. If BBHs with different spin-direction behavior have different $q$ distributions, we can allow for both large $\beta_{q}$, which is typical for dynamical environments, and small $\beta_{q}$, which is typical for highly efficient CE phases, for example. We observe this in Fig.~\ref{fig:beta_IA} where isotropic binaries could mostly prefer equal mass mergers, while aligned binaries also allow for smaller mass ratios. However, Fig.~\ref{fig:beta_IA} shows that the data disfavor the parameter space where both isotropic and aligned channels strongly prefer $q=1$.

To further study this, we fix $\beta_{q,I} = 3, 4, 5$ as a proxy for the mass ratios expected in the dynamical scenario. We plot the recovered $\beta_{q,A}$ in Fig.~\ref{fig:betaI_fixed}. We find the distribution $\beta_{q, A}$ is not very sensitive to the given values of fixed $\beta_{q,I}$. For $\beta_{q,I}=3$, $4$ and $5$ the $90\%$ upper limit on  $\beta_{q,A}$  is $2.25$, $1.85$ and $1.70$ respectively. 
In other words, in these scenarios with a strong mass pairing as expected for GCs, the aligned channel must have a weaker mass pairing. The results of detailed modeling can be mapped onto these findings, as exemplified in Figs.~\ref{fig:beta_IA} and \ref{fig:betaI_fixed} using the CE and GC models of \citet{Zevin:2020gbd}.

To conclude, here we illustrate how differences in isotropic and isolated subpopulations can be leveraged to gain insights into astrophysical processes that govern the merger of BBHs.
We caution that in this section's illustration we only employed detailed modeling from only two formation channels (one each for isotropic and aligned spin binaries) out of a slew of scenarios proposed for synthesizing BBHs. These proposed formation channels are plagued by many astrophysical uncertainties, affecting population properties or merging binaries in highly degenerate ways. Consequently, including only a subset of formation channels can lead to biased inferences on astrophysical uncertainties \citep{Zevin:2020gbd}. \new{
Populations that seem unlikely when evaluated alone can become plausible when multiple channels are taken into account. For instance, aligned-spin binaries formed through chemically homogeneous evolution (CHE) tend to have similar masses and may not seem likely if they coexist alongside isotropic spin binaries assembled in stellar clusters. But, if other channels exist that result in unequal-mass mergers, such as highly efficient CEs, a portion of CHE binaries can also exist. This would allow for a combined aligned-spin population whose mass-ratio distribution is characterized by a small $\beta_{q,A}$.}
While it is not easy to disentangle formation scenarios with similar spin-direction predictions, we can at least break degeneracies between populations that predict isotropic and aligned spins. Using spin directions as an anchor allows us to make more robust claims than those based on astrophysical models alone.

\subsection{Caveats}
\label{sec:caveats}
In this study, we used BH spin tilts to indicate the formation environment: isotropic spins pointing towards the dynamical assembly of BHs and aligned spins implying an isolated origin. However, GWs contain little information about the individual BH spins; hence, they are very poorly measured. Gravitational-wave signals instead depend primarily on the effective spin parameter $\chi_{\rm eff}$ (defined as the BH spin contribution onto the binary's orbital angular momentum) at the leading 2PN order~\citep{Damour:2001tu, Racine:2008qv}.
The only scenarios where individual spin components have been well measured are when the binary is perfectly aligned, when mass ratios are small or when the binary is observed close to edge-on~\citep{Vitale:2016avz} or when the binary is detected with large signal-to-noise ratios \citep{Purrer:2015nkh}. Since the spin tilts are poorly measured, the estimation of hyperparameters that govern the properties of isotropic and aligned populations are also not well measured and depend significantly on the model employed to describe them~\citep{Tong:2022iws, Galaudage:2021rkt,Vitale:2022dpa, Callister:2022qwb}. Hence, the distribution of $\beta_{q, I}$ and $\beta_{q, A}$ recovered in this study are not well measured.

In addition, the mixture model used in this study (as well as the \texttt{Default} spin model in \citet{LIGOScientific:2021psn}) are plagued by the correlation between $\zeta$ and $\sigma_t$.  If the distribution of aligned-spin tilts (parameterized by $\sigma_t$) is constrained to smaller angles, then the inferred branching ratio $\zeta$ is also small, requiring a larger population of isotropic binaries. On the other hand, if $\sigma_t$ is left unconstrained, it shows a tendency to identify even isotropic binaries as aligned by possessing large $\sigma_t$.

Since $\chi_{\rm eff}$ is measured more accurately than the spin directions $\theta_{1,2}$, it can also be used to differentiate between binaries formed in the isolated channel and those assembled dynamically. In this case, the isotropic binaries will possess a symmetric $\chi_{\rm eff}$ distribution centered at $\chi_{\rm eff}=0$, while the preferentially-aligned binaries will have  a distribution skewed towards positive values of $\chi_{\rm eff}$ ~\citep{Rodriguez:2016vmx, Farr:2017gtv, Wysocki:2017isg, Gerosa:2018wbw, Ng:2018neg}. If we use the mixture model described in Eq.~(\ref{eq:mix_pop2}), we find that there is no evidence of multiple populations.  We find that the observations can be explained with a single distribution consistent with the \texttt{Gaussian} model employed in \citet{LIGOScientific:2021psn}. This is consistent with results by \citet{Callister:2021fpo}. However, it is possible that as the number of detections increases with technological improvement in GW detectors, one could discern the differences in isotropic and aligned spin populations using the $\chi_{\rm eff}$ distribution as well.

\new{In this article, we have only used parametric models that extend the \texttt{Default} spin model to isotropic and aligned spin populations to illustrate how spin orientation assumptions can anchor astrophysical inferences. However, simple parametric models make strong assumptions about the underlying distribution and can be biased if the model is not accurate. Studies by \citet{Vitale:2022dpa} and \citet{Edelman:2022ydv} have revealed features in the tilt distribution that can not be accounted for by the \texttt{Default} spin model or our mixture models. We describe spin magnitudes in App.~\ref{sec:sz} using a beta distribution that assumes there are no non-spinning black holes to eliminate singularity at $\chi_{1,2}=0$. However, \citet{Edelman:2022ydv} and \citet{Golomb:2022bon} have demonstrated that flexible, data-driven models provide much greater support for small spins than that allowed by the widely-used beta distribution. Parametric models have also led to conflicting conclusions (e.g. ~\citet{Roulet:2021hcu, Galaudage:2021rkt,Tong:2022iws, Callister:2022qwb,  Vitale:2022dpa}), which highlights the sensitivity of inferences to modeling choices. Mixture models used in this study could also be vulnerable to these inaccuracies and may not fully capture all the details from the catalog, and they need to be tested against data-driven models for more robust conclusions.}

Finally, the model used in this work (and the \texttt{Default} spin model in \citet{LIGOScientific:2021psn}) assumes that these are the only two categories of spin-direction distributions: isotropic binaries originating in the field and aligned binaries originating in dense stellar clusters. \new{However, the spin axis of BHs evolving in isolation can change direction during the core collapse of a star, potentially resulting in an isotropic spin distribution for the first-born BH \citep{Farr:2011gs, Tauris:2022ggv}. Moreover,} formation scenarios -- such as BBHs synthesized in AGN disks -- might predict binary where one or both BH spins are antialigned with orbital angular momentum~\citep{McKernan:2019beu,Tagawa:2020dxe}. We have not considered such populations with preferentially antialigned spins. Since current observations indicate a dearth of binaries with $\cos\theta_i = -1 $ ~\citep{LIGOScientific:2021psn, Galaudage:2021rkt, Callister:2022qwb, Vitale:2022dpa}, this should not significantly affect our results. But if future observations uncover more features in the spin distribution of BBHs, our model can be extended to include an anti-aligned spin component.

\section{Conclusions}
\label{sec:conclusions}

The spin orientations in BH mergers are possibly the cleanest observables to shed light on BH binary formation: BHs assembled during dynamical encounters are expected to be isotropic spins, while those formed in isolation should have spins preferentially aligned with the orbital angular momentum. The approach to distinguish isotropic and preferentially-aligned binaries has already been implemented by  \citet{LIGOScientific:2021psn}. However, the conventional model does not provide any astrophysical insights into the two subpopulations other than their relative abundance or the tilts distribution of the aligned population (which can shine a light on the supernova kicks that misalign the BH spins or other processes that realign them~\citep{Gerosa:2018wbw}). 

In this work, we extend the \texttt{Default} spin model of \citet{LIGOScientific:2021psn}, and use spin tilts as an anchor to extract more information about the distribution of binary properties in the isotropic-spin and aligned-spin populations. %
We find evidence that BHs coming from the two subpopulations have opposite tendencies when forming a pair: if BHs with isotropic spins strongly prefer partners with similar masses, then BHs with aligned spins should be less picky, and vice versa. We discuss what implications this has on the relative abundance of isotropic and aligned binaries and their various correlations with the tilt distribution of the aligned population.
We also demonstrate how differentiating binaries by spin alignment on the population level can provide insights into the unknown physics and various processes that drive BHs to merge. For illustration, we compare our results with the mass-pairing function of BH mergers in globular clusters and those driven by common-envelope (as presented in \citet{Zevin:2020gbd}). We find that the mass-pairing for isotropic-spin BBHs are consistent with the extremely selective pairing expected from globular clusters. In addition, our model allows us to put constraints on astrophysical parameters, such as the common-envelope efficiency. However, such inferences that employ various astrophysical models significantly depend on the underlying uncertainties and could be degenerate with other formation channels not considered in this study. Hence,  it is hard to disentangle individual subpopulations that make up the isotropic/aligned populations. However, using the prescription outlined in this study, we can still make broader claims about the relationship between the mass-ratio distribution of {\em overall} isotropic or aligned populations.

We extend our analysis to distributions of spin magnitudes and redshifts in isotropic and aligned subpopulations in Appendix~\ref{sec:sz}.
However, the current observations are insufficient in discerning the differences in spin magnitude or redshift distribution of isotropic and aligned spin binaries. This could be due to the fact that individual spin tilts are poorly measured through GWs. 
As the current GW detectors undergo further improvements, they will be able to observe a larger number of sources, enabling us to put tighter constraints on the contributions of field and dynamical formation channels to the binary black hole population.

\section*{Acknowledgements}

We are grateful to Salvatore Vitale and Sylvia Biscoveanu for the helpful discussions. This material is based upon work supported by NSF's LIGO Laboratory which is a major facility fully funded by the National Science Foundation.  Z.D. acknowledges support from the CIERA Board of Visitors Research Professorship and NSF grant PHY-2207945. V.K. was partially supported through a CIFAR Senior Fellowship, a Guggenheim Fellowship, the Gordon and Betty Moore Foundation (grant award GBMF8477), and from Northwestern University, including the Daniel I. Linzer Distinguished University Professorship fund. This research was supported in part through the computational resources
and staff contributions provided for the Quest high performance computing facility at Northwestern University which is jointly supported by the Office of the Provost, the Office for Research, and Northwestern University Information Technology.

\bibliography{ref}{}
\bibliographystyle{aasjournal}

\appendix

\section{Hierarchical Bayesian Analysis}
\label{sec:HBA}

We employ a hierarchical Bayesian inference framework to measure the mass, spin and redshift distributions of BBH mergers using only gravitational-wave data. We can parameterize the distribution of the individual binary properties, \vt (like masses, spins, redshifts) in terms of {\em unknown} population-level hyper-parameters \vl. We wish to  infer this \vl given the catalog \vd ($\equiv \{d_i\}$) consisting of the $N_\mathrm{obs}=69$ BBHs reported in GWTC-3 with false alarm ratio smaller than 1 per year~\citep{LIGOScientific:2021psn}.
The posterior of the hyperparameters \vl governing the distributions of \vt, ~\citep{Mandel:2018mve,Fishbach:2018edt,Vitale:2020aaz} is

\be\label{eq:HyperPostOfLikeFirst}
p(\vl | \vd) \propto {\pi(\vl)} \prod_{i=1}^{N_{\rm obs}} \frac{p(d_i |\vl) }{\xi(\vl)}\,.  %
\ee
where $\pi(\vl)$ is the population prior and $p(d_i |\vl) $ is the likelihood of individual BBHs. Since we are only interested in the \textit{shape} of \vt distribution, we have marginalized the overall merger rate density assuming a logarithmically-uniform prior $p(N_\mathrm{obs})\propto N_\mathrm{obs}^{-1}$.

For an individual event, the $p(d_i |\vl)$ can be expressed as
\begin{align}
p(d_i |\vl) =&\int \dd \vt \ p(d_i |\vt )\  p(\vt|\vl)\nn
= \int \dd \vt\; \frac{ p_{\rm pe}(\vt | d_i)p(\vt|\vl) }{p_{\rm pe}(\vt)},\nn\\
 \simeq&  \Bigl<\frac{p(\vt_{i}|\vl) }{p_{\rm pe}(\vt_{i})} \Bigr>  \nn
\end{align}
where $p_{\rm pe}(\vt | d_i)$ is the posterior distribution for the event; $\pi_{\rm pe}(\vt)$ is the original prior adopted during the individual-event parameter estimation; and $p(\vt|\vl)$ is the population model for how the hyperparameters \vl govern the distribution of binary parameters \vt. 
Since we do not have $p_{\rm pe}(\vt | d_i)$ for each event, but discrete samples drawn from  $p_{\rm pe}(\vt | d_i)$, in the last line, we recast the integral as Monte Carlo average over the posterior samples.

Detection efficiency $\xi(\vl)$ in the denominator of Eq.~(\ref{eq:HyperPostOfLikeFirst}) is the fraction of detectable BBHs given the proposed population \vl. This correction accounts for observational selection bias and is given by,
  \begin{equation}
  \xi(\vl) = \int p_\mathrm{det}(\vt) p(\vt|\vl) d\vt,
  \end{equation}
where $p_\mathrm{det}(\lambda)$ is the probability that an event with properties \vt can be recovered by detection pipelines with an FAR$<1\,\mathrm{yr}^{-1}$. 

We can also calculate $\xi(\vl)$ by summing over  $N_\mathrm{det}$ detectable injections out of $N_\mathrm{inj}$ signals drawn from some reference distribution $p_\mathrm{inj}(\vt)$,
    \be
    \xi(\vl) = \frac{1}{N_\mathrm{inj}} \sum_{i=1}^{N_\mathrm{det}} \frac{p(\vt|\vl)}{p_\mathrm{inj}(\vt)}\,,
    \label{eq:xi-mc}
    \ee

We use the \textsc{GWPopulation}~\citep{Talbot:2019okv}
for applying the hierarchical bayesian framework, and we use the \textsc{emcee} sampler (implemented in \textsc{bilby}~\citep{Ashton:2018jfp}) to draw samples from the posterior on $\vl$.

We use parameter estimation samples released LVK: \texttt{Overall\_posterior} parameter estimation samples for BBHs  detected in GWTC-1;
\texttt{PrecessingSpinIMRHM} samples  for events published in GWTC-2 and GWTC-2.1~\citep{LIGOScientific:2020ibl}; and \texttt{C01:Mixed} for events in GWTC-3~\citep{LIGOScientific:2021djp}.\footnote{
Parameter estimation samples are available at\\
GWTC-1: https://dcc.ligo.org/LIGO-P1800370/public\\
GWTC-2: https://dcc.ligo.org/LIGO-P2000223/public\\
GWTC-2.1: https://zenodo.org/record/5117703\\
GWTC-3: https://zenodo.org/record/5546663
}.

We evaluate the detection efficiency $\xi(\vl)$ using successfully recovered BBH injections, released by LVK in \texttt{o1+o2+o3\_bbhpop\_real+semianalytic-LIGO-T2100377-v2.hdf5}\footnote{https://zenodo.org/record/5636816}.

\begin{figure*}
    \includegraphics[width=0.8\textwidth]{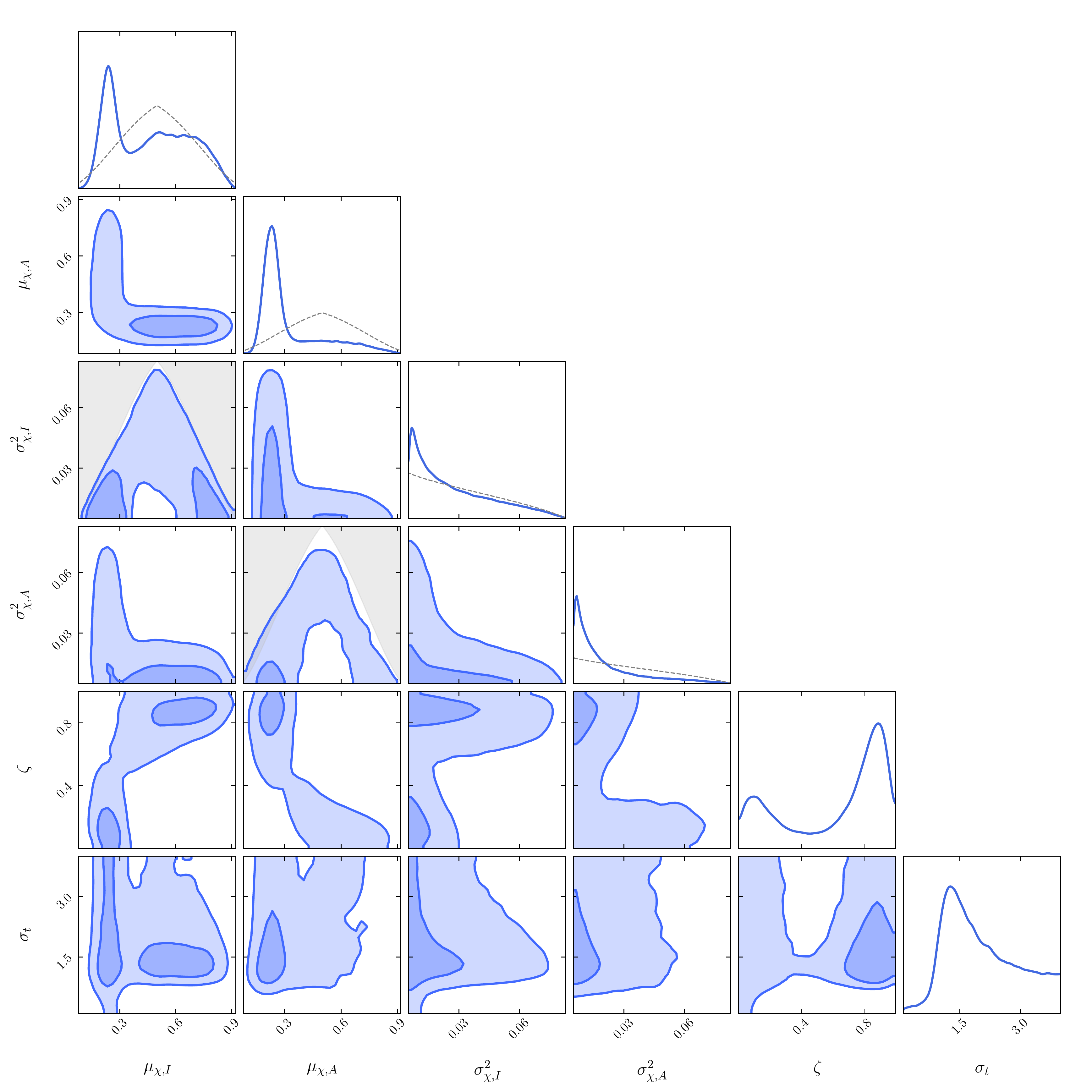}
    \caption{ Posteriors of all the hyperparameters in the \texttt{s-mix} model where ($\mu_{\chi,I}$,  $\mu_{\chi, A}$, $\sigma_{\chi,I}^2$, $\sigma_{\chi, A}^2$ fit the $\vt_{\rm mix}=\{\chi_1, \chi_2 \}$, while $\zeta$ and $\sigma_t$ fit mixing fraction and tilts distribution of aligned systems. We show the $68\%$ and $90\%$ intervals. The grey region marks the parameter space in $(\mu_{\chi, I}, \sigma_{\chi, I}^2)$ and $(\mu_{\chi, A}, \sigma_{\chi, A}^2)$  that is excluded to avoid the singularity of the Beta distribution [Eq.~(\ref{eq:muvar})]. The grey dotted lines are the resulting effective priors $p_{\rm eff}(\mu_\chi)$ and $p_{\rm eff}(\sigma_\chi^2)$ [Eq.~(\ref{eq:peffmu}) and Eq.~(\ref{eq:peffvar}) respectively].}
    \label{fig:s-mix}
\end{figure*}

\section{Spin and Redshift distributions}
\label{sec:sz}

In the previous sections, we analyzed how the pairing function differs between the isotropic and aligned systems. This section will focus on the differences in their spin magnitude and redshift distributions.

Spin magnitudes could be similar for both field binaries and cluster binaries if they are only controlled by stellar collapse physics~\citep{OConnor:2010moj}. However, during the isolated evolution of binaries, processes like stable mass transfer and tidal interactions can spin up the BHs~\citep{Bavera:2020uch, duBuisson:2020asn}. 
In dense stellar clusters, BHs born from previous mergers (instead of stellar collapse) have a characteristic spin $\sim 0.7$ \citep{Gerosa:2017kvu,Fishbach:2017dwv,Kovetz:2018vly}. If these BHs merge repeatedly, a small fraction of BBHs originating from clusters could have larger spins. These differences in spin magnitudes of field and cluster binaries, could be observed in the population of detected BBHs. To analyse if this is the case, we will allow the aligned and isotropic binaries to have different spin-magnitude distributions i.e. $\vt_{\rm mix}=\{\chi_1, \chi_2\}$ governed by hyperparameters $\vl_{\rm mix}=\{\mu_{\chi, I}, \sigma^2_{\chi, I}, \mu_{\chi, A}, \sigma^2_{\chi, A}\}$. We call this model \texttt{s-mix}. Here, we assume component spin magnitudes $\chi_{1,2}$ are identically distributed according to a Beta distribution~\citep{Wysocki:2018mpo} (we have dropped subscripts ``I'' and ``A'' for brevity),
\be
p(\chi_i | \alpha, \beta) = \frac{\chi_i^{1-\alpha} \, (1-\chi_i)^{1-\beta}}{c(\alpha, \beta)}\,,
\label{eq:beta}
\ee
where $c(\alpha, \beta)= \frac {\Gamma (\alpha )\Gamma (\beta )}{\Gamma (\alpha +\beta )}$ is a normalization constant, where $\Gamma$ is the gamma function. We sample the the mean and variance ($\mu_{\chi}, \sigma^2_\chi$)  of the beta distribution (and not in $(\alpha, \,\beta)$ directly)
\begin{align}
  \mu_\chi &= \frac{\alpha_\chi}{\alpha_\chi + \beta_\chi}\\
  \sigma_\chi^{2} &= \frac{\alpha_\chi\beta_\chi}{(\alpha_\chi+\beta_\chi)^{2}(\alpha_\chi + \beta _\chi+ 1)}.
  \label{eq:alphabeta-to-musigma}
\end{align}
The shape parameters $\alpha$ and $\beta$ can be recovered from $\mu_\chi$ and $\sigma^2_\chi$ through
\begin{align}
\alpha_\chi = &  \frac{\mu_\chi^2(1-\mu_\chi)}{\sigma_\chi^2} - \mu_\chi,\nn\\
\beta_\chi  = &  \frac{\mu_\chi(1-\mu_\chi)^2}{\sigma_\chi^2} - (1-\mu_\chi)
\label{eq:musigma-to-alphabeta}
\end{align}

It also implies that there are certain constraints on the distribution's mean and variance $(\mu_\chi, \sigma^2_\chi)$ (which is what we actually sample).
To ensure $\alpha_\chi>1,\, \beta_\chi>1$
\be\label{eq:muvar}
\sigma_{\chi}^2<
\begin{dcases}
\frac{\mu_{\chi} ^2 (a_{\rm max}-\mu_{\chi} )}{(a_{\rm max}+\mu_{\chi} )} & 0<\mu_{\chi} <\frac{a_{\rm max}}{2} \\ \frac{\mu_{\chi}  (a_{\rm max}-\mu_{\chi} )^2}{(2 a_{\rm max}-\mu_{\chi})} & \frac{a_{\rm max}}{2}<\mu_{\chi} <a_{\rm max}
\end{dcases}
\ee
and the maximum variance that can be achieved is $a_{\rm max}^2/12$ at $\mu=a_{\rm max}/2$. %
We will assume $a_{\rm max}=1$ and that the priors on $\mu_{\chi, I}$ and $\mu_{\chi, A}$ are uniform between $0$ and $1$, while the prior on $\sigma^2_{\chi, I}$ and $\sigma^2_{\chi, A}$ are uniform between $0$ and $1/12$.  However, since we have imposed that the Beta distribution is not singular, a vast region in the prior space is restricted. In this case, the effective one-dimensional prior on $\mu_\chi$ is 
\be\label{eq:peffmu}
p_{\rm eff}(\mu_\chi) =\frac{1}{c}
\begin{dcases}
\frac{\mu_{\chi} ^2 (1-\mu_{\chi} )}{(1+\mu_{\chi} )} & 0<\mu_{\chi} <\frac{1}{2} \\ \frac{\mu_{\chi}  (1-\mu_{\chi} )^2}{(2-\mu_{\chi})} & \frac{1}{2}<\mu_{\chi} <1
\end{dcases}
\ee
where $c=4 \log({3}/{2})-{19}/{12}$ is the normalization constant, while the effective one dimensional prior on $\sigma_\chi^2$ is 
\be\label{eq:peffvar}
p_{\rm eff}(\sigma_{\chi}^2) \propto 1+2 \Re\left((1+i \sqrt{3}) \mathcal{S}_{\chi}^{1/3}\right)
\ee

where
\be
\mathcal{S}_{\chi}= 1+3 \sigma_{\chi}  \left(\sqrt{3(\sigma_{\chi}^4+11 \sigma_{\chi}^2-1)}-6 \sigma_{\chi} \right)\,.\nn
\ee

 Fig.~\ref{fig:s-mix} shows results when analyzing GWTC-3 binary BHs with the \texttt{s-mix}.  We plot the $68\%$ and $90\%$ intervals for the mean and variance of $\chi_{1,2}$ distribution for isotropic and aligned populations ($\mu_{\chi,I}, \sigma^2_{\chi, I}$, $\mu_{\chi,A}, \sigma^2_{\chi, A}$), the fraction of aligned binaries ($\zeta$) and standard deviation of aligned-spin tilts $\sigma_t$. We also show the effective priors $p_{\rm eff}(\mu_\chi)$ and $p_{\rm eff}(\sigma_\chi^2)$ [Eq.~(\ref{eq:peffmu}) and Eq.~(\ref{eq:peffvar}) respectively] as well as the region excluded to avoid the singularity of the Beta distribution [Eq.~(\ref{eq:muvar})].  We see the presence of multiple modes in Fig.~\ref{fig:s-mix}. However, these results are consistent with the distribution of spin magnitudes obtained by \texttt{Default} spin model in \citet{LIGOScientific:2021psn}.
We observe two modes in the mixing fraction, $\zeta\simeq 0.1$ (implying that  majority of systems are isotropic) and $\zeta\simeq 0.9$ (implying the majority of systems are aligned). 
When aligned systems are favored, we find that their spin-magnitude distribution peaks at $\mu_{\chi, A}\simeq 0.24$. This is consistent with the \texttt{Default} spin model in \citet{LIGOScientific:2021psn}. Also, the spin magnitudes of isotropic binaries are consistent with the prior when $\zeta$ is large. On the contrary, when isotropic systems are favored (small $\zeta$), isotropic binaries reproduce the \texttt{Default} spin model while the aligned binaries extract the prior. In both cases, we find that the overall model is effectively trying to reproduce the \texttt{Default} model.

For simplicity, in the above model (\texttt{s-mix}), we only fit $\vl_{\rm mix}=\{\mu_{\chi, I}, \sigma^2_{\chi, I}, \mu_{\chi, A}, \sigma^2_{\chi, A}\}$ and $\zeta, \sigma_t$. We keep all the other hyperparameters associated with describing the mass and redshift distributions fixed to their median values obtained in \citet{LIGOScientific:2021psn}. However, \citet{LIGOScientific:2021psn, Callister:2021fpo} have shown that the mass ratios and spins of detected BBHs show evidence of anti-correlation. Since $\chi_{\rm eff}$ depends on the spin tilts, this would require that distribution of spins be fitted alongside the mass-ratio distribution. We use two approaches to test if fitting $q$ distribution alongside spins affects our results. In the first approach, we assume isotropic and aligned binaries have different spin-magnitude distributions (governed by $\vl_{\rm mix}=\{\mu_{\chi, I}, \sigma^2_{\chi, I}, \mu_{\chi, A}, \sigma^2_{\chi, A}\}$) but the same mass ratio distribution (governed by $\beta_q$). This model yields a result similar to the \texttt{s-mix} model with $\beta_q$ consistent with \texttt{Powerlaw+Peak} model  in \citet{LIGOScientific:2021psn}. In the second approach, we assume isotropic and aligned binaries have both different spin-magnitude distributions (governed by $\{\mu_{\chi, I}, \sigma^2_{\chi, I}, \mu_{\chi, A}, \sigma^2_{\chi, A}\}$) and different mass ratio distribution (governed by $\{\beta_{q, I}, \beta_{q, A}\}$). In this case, we find that the spin magnitudes and tilts reproduce the \texttt{s-mix} model with bimodal features. However, $\{\beta_{q, I}, \beta_{q, A}\}$ do not follow the anti-correlation observed in the \texttt{q-mix} model. Instead,  $\{\beta_{q, I}, \beta_{q, A}\}$ also show bimodality similar to spin magnitudes i.e. dominant channel recovers the LVK result while the subdominant channel is consistent with prior. This is because assuming different spin-magnitude distributions for aligned and isotropic binaries enforces that either channel dominates and reproduces the LVK result. In Sec.~\ref{sec:q-dist}, we discussed that features in $\beta_{q, I}$ and $\beta_{q, A}$, observed with \texttt{q-mix} model, could help explain the $q-\chi_{\rm eff}$ correlation observed in GWTC-3~\citep{LIGOScientific:2021psn}. If that were the case, one would also expect that isotropic and aligned subpopulations have different spin magnitudes. This could be understood through following toy model. Let's assume that BBHs with aligned (isotropic) binaries have spin magnitudes $\chi_A$ ($\chi_I$). Also the \texttt{q-mix} model is consistent with aligned (isotropic) binaries  having $q\simeq 1$ (small mass ratios). In that case, the $\chi_{\rm eff}$ distribution of isotropic binaries should be
\be
\chi_{\rm eff, I}= \chi_I \frac{\cos\theta_1+\cos\theta_2}{2}
\ee
This yields a triangle distribution (see~\citet{Baibhav:2020xdf} for derivation) of $\chi_{\rm eff, I}$ between $-\chi_{I}$ and $\chi_{ I}$. On the other hand,  $\chi_{\rm eff}$ distribution of aligned binaries (assuming small mass ratios) should be
\be
\chi_{\rm eff, A} \simeq \chi_A \cos\theta_1
\ee
This would yield $\chi_{\rm eff}$ distribution symmetric around $0$ at $q\simeq1$, and positive values with peak at $\chi_{\rm eff}=\chi_A$ at small $q$. However, since the current data can not discern differences in the spin-magnitude distribution of isotropic and aligned binaries, we can not test our hypothesis that $q-\chi_{\rm eff}$ correlation is caused by isotropic and aligned binaries having different properties. However, as more information is available with the next observing runs of LVK, it might be possible to gain insight into the cause behind such features.

The formation history of BBHs in most channels is predicted to be set by the star formation history and hence, could be similar across binaries originating in galactic fields and stellar clusters. However,  different channels have different processes that drive the merger. These processes are not equally efficient and might have different time delays between binary formation and merger~\citep{vanSon:2021zpk, Mapelli:2018wys, Baibhav:2019gxm}. For example, even in the field, binaries mergers driven by stable mass transfer take significantly longer than those driven by CE. These differences might be observable in the redshift distribution of BBHs detected by LVK~\citep{Fishbach:2021mhp}. To analyze these differences, we allow the aligned and isotropic binaries to have redshift distributions, i.e., $\vt_{\rm mix}=\{z\}$ governed by hyperparameters $\vl_{\rm mix}=\{\kappa_{ I},  \kappa_{ A}\}$. We call this model \texttt{z-mix}. Here $\kappa_{I, A}$ is the power-law slope that governs the evolution of the source-frame merger rate,
\be
p(z|\kappa_{I, A}) \propto \frac{1}{1+z}\frac{dV_c}{dz} \left(1+z\right)^{\kappa_{I, A}}\,,
\ee
where $\frac{dV_c}{dz}$ is the differential comoving volume per unit redshift. We assume an uniform prior on $\kappa_{I}$ and $\kappa_A$ between $-10$ and $10$. We find that \texttt{z-mix} model can not distinguish between the redshift distribution of isotropic and aligned systems. We find that the $\zeta$ distribution is consistent with flat prior (unlike \texttt{s-mix} where $\zeta$ was bimodal). In addition, we find that  either $\kappa_I$ or $\kappa_A$  recover $\kappa$ calculated in \citet{LIGOScientific:2021psn}.

\end{document}